%% file: icdcs25-s2m3.tex
\documentclass[conference]{IEEEtran}
\IEEEoverridecommandlockouts
\usepackage{cite}
\usepackage{amsmath,amssymb,amsfonts}
\usepackage{algorithmic}
\usepackage{graphicx}
\usepackage{textcomp}
\usepackage{xcolor}
\def\BibTeX{{\rm B\kern-.05em{\sc i\kern-.025em b}\kern-.08em
    T\kern-.1667em\lower.7ex\hbox{E}\kern-.125emX}}

\usepackage{subfigure}
\usepackage{capt-of}
\usepackage{booktabs}
\usepackage{varwidth}
\usepackage{graphicx}
\usepackage{kotex}
\usepackage{xcolor}         
\usepackage{graphicx}
\usepackage{subfigure}
\usepackage{multirow}
\usepackage{multicol}
\usepackage{algorithm}
\usepackage{algorithmic}
\usepackage{sidecap}
\usepackage{color}
\usepackage{colortbl}
\usepackage{diagbox}
\usepackage{amsmath}
\usepackage{enumitem}
\usepackage{caption}
\usepackage{nicematrix}
\usepackage{enumitem}
\usepackage{bbm}
\newsavebox\tmpbox
\usepackage{hyperref}
\usepackage{pifont}
\usepackage{wrapfig}

\newcommand{\refsec}[1]{Sec\@.~\ref{sec:#1}}
\newcommand{\reffig}[1]{Fig\@.~\ref{fig:#1}}
\newcommand{\reftab}[1]{Table~\ref{tab:#1}}

\newcommand{\refalg}[1]{Algorithm~\ref{alg:#1}}
\newcommand{\refinsight}[1]{Insight~\ref{insight:#1}}

\newcommand{\Revise}[1]{\textcolor{black}{#1}}

\newcommand{\ours}{S2M3}

\newtheorem{insight}{Insight}
\newtheorem{implication}{Remark}

\DeclareMathOperator*{\argmin}{arg\,min}

\begin{document}

\title{\ours: Split-and-Share Multi-Modal Models \\for Distributed Multi-Task Inference on the Edge}

\author{
    \IEEEauthorblockN{JinYi Yoon\IEEEauthorrefmark{1}, JiHo Lee\IEEEauthorrefmark{1}, Ting He\IEEEauthorrefmark{2}, Nakjung Choi\IEEEauthorrefmark{3}, and Bo Ji\IEEEauthorrefmark{1}}
    \IEEEauthorblockA{\IEEEauthorrefmark{1}Virginia Tech, Blacksburg, VA, USA}
    \IEEEauthorblockA{\IEEEauthorrefmark{2}Pennsylvania State University, University Park, PA, USA}
    \IEEEauthorblockA{\IEEEauthorrefmark{3}Nokia Bell Labs, Murray Hill, NJ, USA\\
    \{jinyiyoon, jiholee\}@vt.edu, tinghe@psu.edu, nakjung.choi@nokia-bell-labs.com, boji@vt.edu}
}

\maketitle

\begin{abstract}
\input{0_abstract_v3}
\end{abstract}


\section{Introduction}\label{sec:intro}

\input{1_intro_v3}
\section{Related Work}\label{sec:related}
\input{2_related_v3}
\section{Background}\label{sec:background}
\input{3_background_v3}
\section{Split-and-Share Architecture}\label{sec:design}
\input{4_design_v3}
\section{Placement and Routing in \emph{\ours}}\label{sec:placement}
\input{5_placement_v4}

\section{Experiments}\label{sec:experiments}

\input{6_experiments_v3}
\vspace{-0.1mm}
\section{Conclusion}\label{conclusion}
\vspace{-0.1mm}

\input{7_conclusion_v3}
\vspace{-0.1mm}
\section*{Acknowledgments}

This research was supported in part by NSF grants CNS-2315851 and CNS-2106294, the Commonwealth Cyber Initiative (CCI) of Virginia, and a Virginia Tech Presidential Postdoctoral Fellowship.

\bibliographystyle{IEEEtran}
\bibliography{icdcs25-s2m3}

\end{document}

%% file: 0_abstract_v3.tex
With the advancement of Artificial Intelligence (AI) towards multiple modalities (language, vision, speech, etc.), multi-modal models have increasingly been used across various applications (e.g., visual question answering or image generation/captioning). Despite the success of AI as a service for multi-modal applications, it relies heavily on clouds, which are constrained by bandwidth, latency, privacy concerns, and unavailability under network or server failures. While on-device AI becomes popular, supporting multiple tasks on edge devices imposes significant resource challenges. To address this, we introduce \emph{\ours}, a split-and-share multi-modal architecture for multi-task inference on edge devices. Inspired by the general-purpose nature of multi-modal models, which are composed of multiple modules (encoder, decoder, classifier, etc.), we propose to \emph{split} multi-modal models at functional-level modules; and then \emph{share} common modules to reuse them across tasks, thereby reducing resource usage. To address cross-model dependency arising from module sharing, we propose a greedy \emph{module-level placement} with \emph{per-request parallel routing} by prioritizing compute-intensive modules. Through experiments on a testbed consisting of 14 multi-modal models across 5 tasks and 10 benchmarks, we demonstrate that \emph{\ours} can reduce memory usage by up to 50\% and 62\% in single-task and multi-task settings, respectively, without sacrificing accuracy. Furthermore, \emph{\ours} achieves optimal placement in 89 out of 95 instances (93.7\%) while reducing inference latency by up to 56.9\% on resource-constrained devices, compared to cloud AI.

%% file: 1_intro_v3.tex
As Artificial Intelligence (AI) systems advance towards multiple modalities (language, vision, speech, etc.) for better recognition and interactive services~\cite{huang2021makes}, multi-modal models that emulate human-like multi-modal understanding are gaining great attention. In recent years, AI as a service (AIaaS) such as ChatGPT~\cite{chatgpt} or Gemini~\cite{team2023gemini} has achieved remarkable success for image-text (and more) tasks and revolutionized user experiences, enabling a wide range of applications including Visual Question Answering (VQA)~\cite{liu2023visual}, image generation/captioning~\cite{hu2022expansionnet}, and cross-modal alignment~\cite{girdhar2023imagebind}. While AIaaS benefits from the powerful resources of cloud computing, it faces significant challenges due to its heavy reliance on the remote server. The growing demand for cloud AI creates communication and bandwidth bottlenecks and thereby increased latency. Clients also encounter unavailability issues during network disruptions or server outages. Furthermore, data privacy is a major concern for cloud-based AI services, as they require users to upload raw data including images.

Edge AI has emerged as a promising approach to alleviating the reliance on clouds by deploying models into the area of operations, especially on edge devices near users such as smartphones, laptops, IoT devices, or sensors. While this approach can provide stable services through local processing, deploying AI on a single edge device
still faces severe resource constraints. Meanwhile, multi-modal models are rapidly growing in size, often exceeding the capacity of most edge devices. To address this, resource-efficient techniques such as model compression via pruning~\cite{zhang2023loraprune,ma2023llm,frantar2023sparsegpt}, knowledge distillation~\cite{zhang2023faster,sun2023dime,chu2024mobilevlm,dai2022enabling}, or quantization~\cite{wang2023bitnet} have been introduced to reduce the model size without noticeable loss in accuracy. However, even a compressed large model can still be too large to fit into a single edge device (e.g., a 4-bit quantized version of a 100B-parameter model is still 200GB/4=50GB in size), and too much compression will lower the inference accuracy~\cite{chen2024comprehensive}.

Furthermore, with the growing popularity of AI systems, it is expected to perform various tasks on the edge. For example, smartphones are expected to support intelligent assistants, image searching/retrieval, text extraction/translation from photos, or face/voice authentication. Deploying separate models dedicated to each task leads to high deployment costs (e.g., memory requirements). Interestingly, there is a limited number of popular data modalities, and most services rely on similar functional modules--such as understanding images for many image-related services or generating text. Thus, avoiding redundant modules shared by multiple tasks can prevent excessive resource usage on such devices.

In this work, we revisit multi-modal model deployment in multi-task scenarios while maintaining efficient memory usage, low latency, and high accuracy. In particular, multi-modal models consist of multiple modules (e.g., vision/text encoders, language models, and classifiers), and different tasks often require modules with similar functions that can potentially be shared so as to reduce memory usage. We aim to split models into modules; identify such shareable modules across tasks; and design resource-efficient inference. This module-level architecture makes compatible with other resource-efficient techniques, including model compression, Deep Neural Network (DNN)~\cite{mohammed2020distributed,li2024distributed} partitioning, and Transformers/Large Language Model (LLM) partitioning methods~\cite{borzunov2024distributed,patel2024splitwise,zhang2024edgeshard,shoeybi2019megatron,osawa2023pipefisher,miao2022galvatron}.

\newcommand{\cmark}{\ding{51}}
\newcommand{\xmark}{\ding{55}}
\begin{table*}[!t]
    \centering
    \caption{Comparison to existing approaches for resource-efficient deployment. Lightweight model approaches include small models, compressed models, or dynamic neural networks. Intra-module partitioning is to divide a single model into layer-, neuron-, or block-wise, while our inter-module partitioning is to decompose a multi-modal model into functional-level modules.}
    \label{tab:comparison}
    \begin{tabular}{lllllll}
        \toprule
        \textbf{Approach} & \textbf{Training/Inference} & \textbf{Design} & \textbf{Multi-Modal} & \textbf{Multi-Task} \\
        \midrule
        LoRAPrune~\cite{zhang2023loraprune}, MobileSAM~\cite{zhang2023faster}, BitNet~\cite{wang2023bitnet}, MoEfication~\cite{zhang2021moefication} & Training & Lightweight Model & \xmark & \xmark \\
        VLKD~\cite{dai2022enabling}, MoE-LLAVA~\cite{lin2024moe}, TinyLLaVA~\cite{zhou2024tinyllava}, LLaVA-Phi~\cite{li2023textbooks} & Training & Lightweight Model & \cmark & \xmark \\
        DIME-FM~\cite{sun2023dime}, MobileVLM~\cite{chu2024mobilevlm}, Edge-MoE~\cite{sarkar2023edge}, Uni-MoE~\cite{li2025uni} & Training & Lightweight Model & \cmark & \cmark \\
        Megatron-LM~\cite{shoeybi2019megatron}, PipeFisher~\cite{osawa2023pipefisher}, Galvatron~\cite{miao2022galvatron} & Training & Intra-Module Partitioning & \xmark & \xmark \\
        DistMM~\cite{huang2024distmm}, DistTrain~\cite{zhang2024disttrain}, Optimus~\cite{feng2024optimus} & Training & Intra-Module Partitioning & \cmark & \xmark \\
        \midrule
        LLM-Pruner~\cite{ma2023llm}, SparseGPT~\cite{frantar2023sparsegpt}, LGViT~\cite{xu2023lgvit}, PowerInfer~\cite{xue2024powerinfer} & Inference & Lightweight Model & \xmark & \xmark \\
        DINA~\cite{mohammed2020distributed}, A3C~\cite{li2024distributed}, PETALS~\cite{borzunov2024distributed}, Splitwise~\cite{patel2024splitwise}, EdgeShard~\cite{zhang2024edgeshard} & Inference & Intra-Module Partitioning & \xmark & \xmark \\
        \midrule
        \textbf{\emph{\ours} (Ours)} & Inference & Inter-Module Partitioning & \cmark & \cmark \\
        \bottomrule
    \end{tabular}
    \vspace{-3mm}
\end{table*}

We design \emph{\ours}, a \underline{\textbf{S}}plit-and-\underline{\textbf{S}}hare \underline{\textbf{M}}ulti-\underline{\textbf{M}}odal \underline{\textbf{M}}odel architecture for multi-task inferences over distributed, resource-constrained devices. We offer solutions to addressing two key challenges in terms of memory constraints and latency:

(i) \textit{How to deploy multi-modal multi-task models on resource-constrained edge devices? -- Split-and-share architecture:} 
Considering that resources are typically scattered across devices, we propose splitting and deploying models into functional-level modules: multiple modality-wise encoders and a task-specific head, reducing the resource requirement on a single device. We then enable the module sharing across different models, allowing the reuse of existing modules when introducing new tasks, further reducing total placement costs. Unlike typical standalone AI, where \Revise{multiple separate copies of the same modules} 
are dedicated to each task, our \emph{split-and-share} architecture requires only a limited number of modality-wise and task-specific modules (see details in~\refsec{design}).

(ii) \emph{Where to place and how to route to serve inference requests within reasonable latency? -- Module-level greedy placement and per-request parallel routing:} 
Our \emph{split-and-share} architecture, where modules can be shared across different models, introduces cross-model dependency; although the requests arrive at the model level, we perform the placement and routing at the module level. To minimize the overall inference latency, we propose a greedy \emph{module} placement and routing \Revise{at functional-level} based on module completion time. Our module-level deployment further enables \emph{per-request parallel routing} over different modality-wise encoders and can compensate for the computational constraints of resource-constrained edge devices, resulting in reasonable latency compared to clouds (see details in~\refsec{placement}).

To the best of our knowledge, \emph{\ours} is the first distributed inference framework designed for multi-modal multi-task at the edge. Our contributions can be summarized as follows:
\begin{itemize}
    \item We \emph{split} the multi-modal models at the functional-level modules to reduce the resource requirements while maintaining accuracy. This functional-level splitting makes it flexible for each module to be interchangeable and compatible with modules having the same function of high-performing, compressed, or partitioned versions.
    \item We \emph{share} common functional-level modules across various tasks, reducing total deployment costs. To address cross-model dependencies due to module sharing, we formulate a module-level placement and routing problem, aiming to minimize the inference latency. We also enable per-request parallel routing across different modalities, effectively reducing inference latency.
    \item Through extensive evaluations using 14 models across 5 tasks and 10 benchmarks on edge devices, we demonstrate that \emph{\ours} can save the placement cost by up to 50\% via splitting and 62\% via sharing across multiple tasks while achieving optimal placement in 89 out of 95 instances (93.7\%). Furthermore, our parallel routing reduces the inference latency by up to 56.9\% with only edge devices, compared to centralized cloud processing.
\end{itemize}

%% file: 2_related_v3.tex
Our work lies in enabling multi-modal multi-task inference on resource-constrained devices, as summarized in \reftab{comparison}.

\textbf{Lightweight models.}
To save resources, lightweight models have been developed by constructing small models~\cite{zhou2024tinyllava,li2023textbooks} or compressed models via pruning~\cite{zhang2023loraprune,ma2023llm,frantar2023sparsegpt}, knowledge distillation~\cite{zhang2023faster,sun2023dime,chu2024mobilevlm,dai2022enabling}, or quantization~\cite{wang2023bitnet}. However, even these models may be too large to fit into a single device; and if making models more lightweight to fit into resource constraints, they often suffer from a trade-off of the accuracy drops~\cite{chen2024comprehensive}. Furthermore, most of these approaches are not plug-and-play, requiring post-training/fine-tuning using some target data with additional computation time.

On the other hand, some dynamic neural networks have been studied to selectively activate only partial, sparse neurons to accelerate the training or inference via masking out some neurons~\cite{xue2024powerinfer}, Mixture-of-Expert (MoE)~\cite{li2025uni,lin2024moe,sarkar2023edge,zhang2021moefication}, or early-exiting~\cite{xu2023lgvit}. It enables multiple benchmarks to maintain their accuracy on top of a single model, but there is no actual resource reduction. They rather require additional resources in general to load all of the parameters (and then dynamically sparsify) along with additional benchmark- or task-specific information. Furthermore, all of these approaches still need additional computation to adapt to target tasks.
In contrast to these lightweight models to customize models for each benchmark, we leverage pretrained models without modifying but decomposing the model, thus maintaining accuracy.

\textbf{Distributed architecture.}
As one of the promising approaches to deploy large models on resource-constrained edge devices, distributed deployment in the context of model parallelism or partitioning has been extensively studied for DNNs~\cite{mohammed2020distributed,li2024distributed} and recently Transformers/LLMs~\cite{borzunov2024distributed,patel2024splitwise,zhang2024edgeshard,shoeybi2019megatron,osawa2023pipefisher,miao2022galvatron}. These approaches reduce the resource requirement on a single device by allocating the layer-/neuron-/block-wise submodels across devices. However, these partitioning methods can incur excessive transmission costs to deliver the data back and forth or additional cache memory to store the historical data, especially in regressive networks such as Transformers. Moreover, these traditional techniques have largely focused on intra-module partitioning, designed for uni-modal models, with limited consideration for multi-modality. 

Multi-modal models differ from uni-modal models in their use of multiple functional modules. By leveraging this unique nature, we consider \emph{inter-module partitioning} by dividing one multi-modal model (e.g., CLIP) into multiple functional modules (e.g., vision encoder, text encoder, and distance function). Note that existing intra-module approaches, such as compression or partitioning, are applied within a single model and thus \emph{orthogonal} to our inter-module partitioning approach, which can complement our partitioning further. 

More closely related to our \textbf{multi-modal} settings, there are a few recent works on distributed multi-modal models~\cite{huang2024distmm,zhang2024disttrain,feng2024optimus}, but they are all designed for training, focusing on training acceleration by pipelining multiple training batches (similar to the typical pipelining in uni-modal training scheduling~\cite{narayanan2019pipedream}). In contrast, we enable parallel processing over different modalities even in a single multi-modal inference request, and thus our per-request parallelization differs from existing pipelining approaches over multiple requests. Furthermore, each of them considers only one specific task, while \emph{\ours} is task-agnostic.

Furthermore, existing works on \textbf{multi-task} inference require a separate model for every single task and have not considered any further resource saving across different multi-modal tasks, incurring an excessive cost to deploy a dedicated model for each task. Interestingly, various pretrained multi-modal models share identical modules that are referred to as \emph{modules} in this paper (e.g., encoders and language models), which have the same architecture and parameters--which can possibly be shared across multi-task models. However, no prior work has explored splitting multi-modal models into functional-level and reusing the common modules of multi-modal models to enable resource-efficient multi-task deployment. 

%% file: 3_background_v3.tex
In this section, we aim to discuss \emph{the unique properties of multi-modal models in multi-task scenarios}.

When processing on-device multi-modal tasks, it is impractical to train from scratch. Edge devices are often constrained in accessing various data necessary to construct robust models. Furthermore, even with enough training data, edge devices suffer from computational or power constraints, often lacking resources. Without powerful resources such as GPUs, the device can take tens to hundreds of minutes (e.g., 89~minutes on the Jetson Nano device~\cite{nvidia_jetson_nano} to train one epoch on the entire Food-101~\cite{bossard2014food} training dataset using ResNet50~\cite{he2016deep}). It also requires high storage capacity to save training data (5GB for Food-101 training dataset) on devices. Fortunately, training from scratch is no longer necessary. Platforms like Hugging Face~\cite{huggingface} provide various pretrained models including Foundation Models (FMs) pretrained on massive datasets, and devices can simply download and just provide fast inference.

\vspace{-0.5mm}
\subsection{Multi-Modal FMs}
\vspace{-0.5mm}

Multi-modal FMs are used for diverse tasks: 1) image-text retrieval: to retrieve relevant text based on an image or vice versa; 2) VQA: to answer questions on a given image; 3) cross-modal alignment: to match visual and textual data to each other (can be extended to align beyond image and text data); 4) image captioning/generation: to generate descriptive texts for an image or vice versa; and others.

\begin{figure}
    \centering
    \vspace{-1mm}
    \subfigure[Encoder VQA, image-text retrieval]{\includegraphics[width=.27\columnwidth]{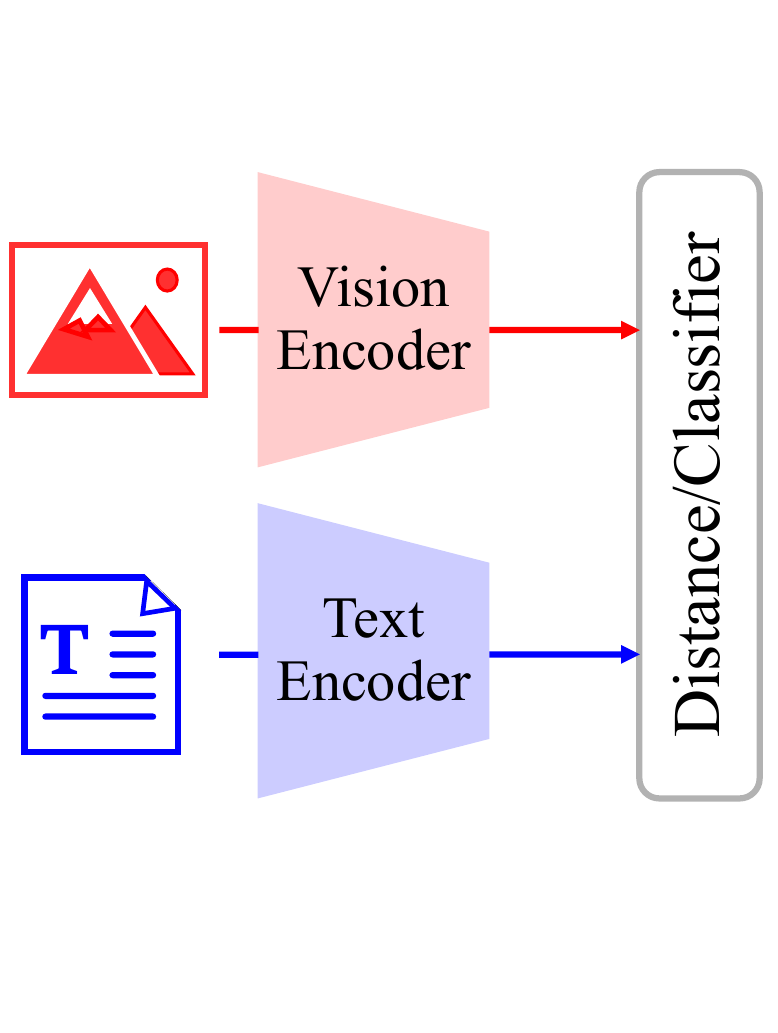}\label{fig:arch-encodervqa}}\hspace{3mm}
    \subfigure[Cross-modal alignment]{\includegraphics[height=.37\columnwidth]{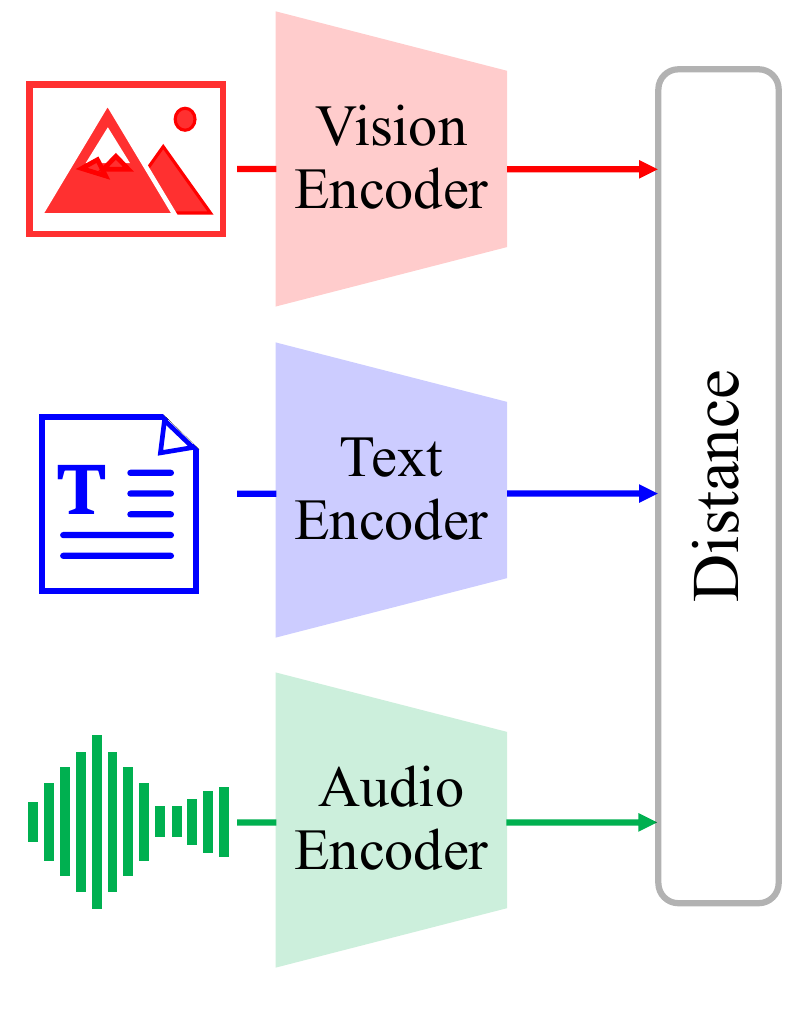}\label{fig:arch-imagebind}}\hspace{3mm}
    \subfigure[Image captioning]{\includegraphics[width=.31\columnwidth]{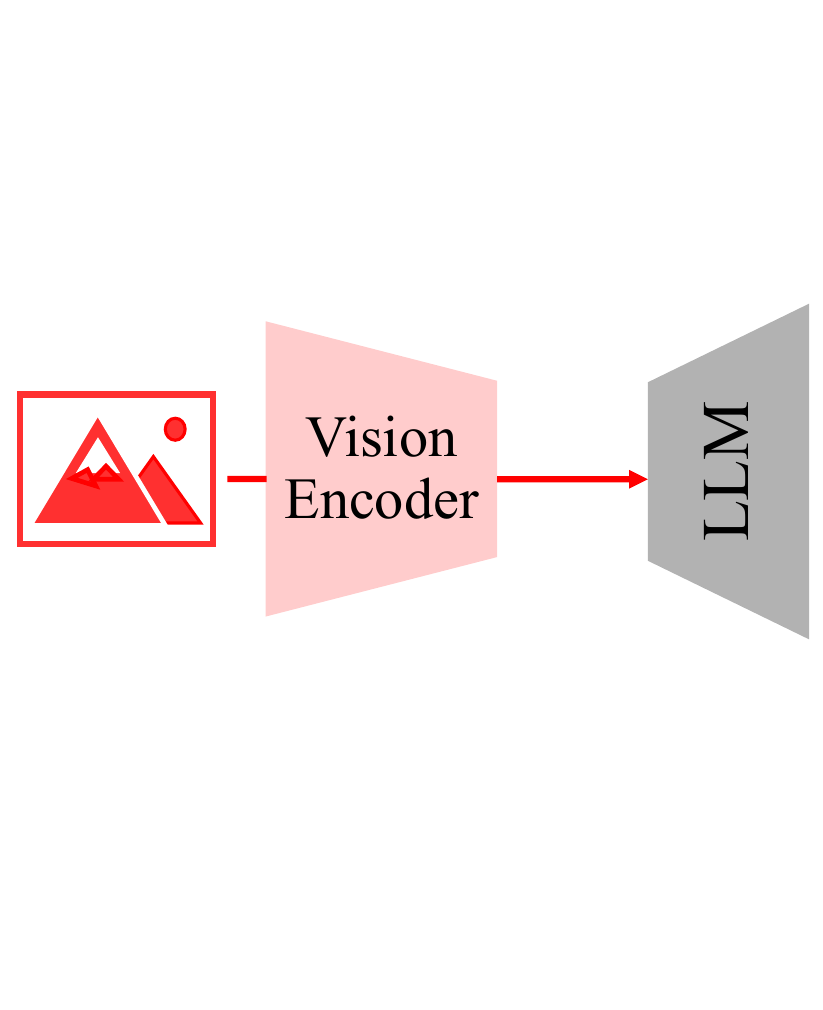}\label{fig:arch-captioning}}
    \vspace{-1mm}
    \caption{Architectures of multi-modal tasks.}
    \vspace{-5mm}
    \label{fig:vl-task-arch}
\end{figure}

We leverage existing FMs to ensure good accuracy on the edge. Here, each multi-modal task requires a specific architecture. As illustrated in \reffig{vl-task-arch}, the image-text retrieval task (\reffig{arch-encodervqa}) has two different input modalities and therefore needs image and text encoders. To identify the most relevant match, some distance measures (e.g., cosine similarity) can be employed as a task head. In cross-modal alignment tasks (\reffig{arch-imagebind}), multiple modalities beyond image or text can also be applied. It has multiple encoders for each modality and then aligns the encoded features using loss functions (e.g., InfoNCE). Similarly, image captioning (\reffig{arch-captioning}) and image classification (with a classifier instead of LLM) also have a similar architecture, which uses an image encoder to understand image data and a task-specific head module to generate outputs relevant to respective tasks.

\vspace{-0.5mm}
\subsection{Key Insights}
\vspace{-0.5mm}

Overall, each task needs encoders for each modality in input data, as well as subsequent models or measures specific to the task. For example, vision-related tasks typically need a vision encoder to understand the input image. Image-text retrieval, encoder-only VQA, and cross-modal alignment tasks also need text encoders to interpret input prompts. After that, they need some distance/similarity measures to evaluate the closeness between modalities. LLM-based VQA and image captioning tasks require text generation to produce text outputs. Although not all models are designed to be split, most multi-modal models commonly have separate encoders for multiple modalities.

\begin{insight}[Splittable architecture]\label{insight:splittable}
    Multi-modal models can be decomposed into function-level modules: 1) multiple modality-wise encoders: to interpret each modality data; and 2) a single task-specific head: to generate task-relevant outputs.
\end{insight}

Given a single inference request, each input data with a different modality is injected into each corresponding encoder and processed independently. There is no data exchange or synchronization during modality encoding in most models.

\begin{insight}[Parallel processing]\label{insight:parallel}
    Module-level splitting enables parallel processing on multiple modality-wise encoders.
\end{insight}

\begin{table}[!t]
    \centering
    \caption{Architectures of multi-modal models.}
    \vspace{-1mm}
    \label{tab:fm-arch}
    \setlength{\tabcolsep}{1.85pt}
    \begin{tabular}{l|ccc|c}
        \toprule
        (Task) & Vision & Text & Audio & Task \\
        \hspace{3mm}Model & Encoder & Encoder & Encoder & Head \\ \midrule

        \multicolumn{1}{l}{(Image-Text Retrieval)} \\
        \hspace{3mm}CLIP ResNet-50 & ResNet-50 & \multirow{9}{*}{TRF} & \multirow{9}{*}{N/A} & \multirow{9}{*}{\shortstack{Cosine\\Similarity}} \\             
        \hspace{3mm}CLIP ResNet-101 & ResNet-101 & & & \\ 
        \hspace{3mm}CLIP ResNet-50x4 & ResNet-50x4 & & & \\ 
        \hspace{3mm}CLIP ResNet-50x16 & ResNet-50x16 & & & \\ 
        \hspace{3mm}CLIP ResNet-50x64 & ResNet-50x64 & & & \\ 
        \hspace{3mm}CLIP ViT-B/32 & ViT-B/32 & & & \\ 
        \hspace{3mm}CLIP ViT-B/16 & ViT-B/16 & & & \\ 
        \hspace{3mm}CLIP ViT-L/14 & ViT-L/14 & & & \\ 
        \hspace{3mm}CLIP ViT-L/14@336 & ViT-L/14@336 & & & \\ \midrule
         
        \multicolumn{1}{l}{(VQA)} \\
        \hspace{3mm}Encoder-only (S) & ViT-L/14@336 & \multirow{2}{*}{TRF} & \multirow{2}{*}{N/A} & \multirow{2}{*}{Classifier} \\
        \hspace{3mm}Encoder-only (L) & ViT-B/16 & & & \\ \cmidrule{1-5} 
        
        \hspace{3mm}LLaVA-v1.5-7B & \multirow{7}{*}{ViT-L/14@336} & \multirow{7}{*}{N/A} & \multirow{7}{*}{N/A} & \multirow{2}{*}{Vicuna-7B} \\
        \hspace{3mm}LLaVA-Next-7B & & & & \\ \cmidrule{1-1} \cmidrule{5-5} 
        \hspace{3mm}LLaVA-v1.5-13B & & & & \multirow{2}{*}{Vicuna-13B} \\ 
        \hspace{3mm}LLaVA-Next-13B & & & & \\ \cmidrule{1-1} \cmidrule{5-5} 
        \hspace{3mm}xtuner-Phi-3-Mini & & & & Phi-3-Mini \\
        \hspace{3mm}Flint-v0.5-1B & & & & TinyLlama-1.1B \\\cmidrule{1-5} 
        \hspace{3mm}LLaVA-v1.5-7B (S) & \multirow{2}{*}{ViT-B/16} & \multirow{2}{*}{N/A} & \multirow{2}{*}{N/A} & Vicuna-7B \\ 
        \hspace{3mm}Flint-v0.5-1B (S) & & & & TinyLlama-1.1B \\ \midrule 
         
        \multicolumn{2}{l}{(Cross-Modal Alignment)} \\
        \hspace{3mm}ImageBind 
        & ViT-H/14 & TRF & ViT-B & InfoNCE \\ \midrule
        
        \multicolumn{1}{l}{(Image Captioning)} \\
        \hspace{3mm}NLP Connect & ViT-B/16 & N/A & N/A & GPT2 \\ \bottomrule
    \end{tabular}
    \vspace{-3mm}
\end{table}

As shown in~\reftab{fm-arch}, for a vision encoder to extract features from input image data, the core functionality is the same, indicating interchangeability. Similarly, language models used for generating answers can be substituted with other language models. For example, FMs for VQA can use Vicuna, Phi-3-Mini, or other language models along with the vision encoder. Moreover, functional modules even have common architectures, where some FMs are built on top of existing ones. For example, Vicuna is finetuned from a model of LLaMA 2. It indicates that modules can be easily swapped to a specific task. 

\begin{insight}[Interchangeability of functional modules]\label{insight:interchange}
    Functional-level split modules offer modular flexibility, allowing for the replacement of individual modules with advanced, compressed, or partitioned versions to adapt to requirements.
\end{insight}

Moreover, many FMs often freeze the modules and have the same functional modules (with the same parameter values) that can be reused, which simplifies the process of building and improving FMs. For example, ViT-B/16 is used in image-text retrieval, VQA, and image captioning tasks. This suggests that if we have one FM with a vision encoder, a text encoder, and a language model, then we can seamlessly reuse these modules in most other multi-modal tasks.

\begin{insight}[Shareable modules]\label{insight:sharable}
    Since functional modules across different tasks/applications share a common architecture, they can be reused and adapted, thereby reducing the time and resources required to deploy new tasks.
\end{insight}

%% file: 4_design_v3.tex
 
We utilize nearby devices for distributed yet cooperative inference. 
\emph{\ours} consists of: 1) \emph{split} architecture to decompose the model into functional modules (in \refsec{distributed}); and \emph{shareable} architecture to reuse modules across tasks (in \refsec{split-and-share}).

\vspace{-0.5mm}
\subsection{Split Architecture in Multi-Modal Inference}
\vspace{-0.5mm}
\label{sec:distributed}

We \emph{plug} pretrained models upon requested tasks and \emph{play} the inference directly without modification, ensuring good accuracy. First, inspired by \refinsight{splittable}, we modularize the model at a functional level and then deploy these modules across devices within resource constraints. Then, following \refinsight{parallel}, we present a distributed inference with parallel processing to compensate for the computational limitations of edge devices.

\textbf{Functional-level modularization.}
We decompose the model into functional modules, specifically 1) modality-wise encoding modules; and 2) a task-specific head module. While some modules can even be further split within the module, we adopt a more coarse approach by dividing the models into functional-level. Thereby, our inter-module (i.e., module-level) partitioning allows pretrained models to be flexible, enabling any intra-module (i.e., layer, neuron, or block-level) approaches of compressed or partitioned versions compatible with \emph{\ours} to boost inference or reduce resource usage.

Based on our split architecture for multi-modal models, we can compute the deployment complexity. 
Let $\mathcal{M}_k := \mathcal{M}^{\mathrm{enc}}_k \cup \{ h_k \}$ be the set of functional modules composing model $k$, where $\mathcal{M}^{\mathrm{enc}}_k$ and $h_k$ are the set of encoder modules and the sole head module of model $k$, respectively. 
For each module $m\in \mathcal{M}_k$, let $r_m$ denote its memory requirement. 
Then, the worst deployment cost on a single device in \emph{\ours} is $\mathrm{max}_{m \in \mathcal{M}_k} r_m$, whereas the deployment cost without split architecture is $\sum_{m \in \mathcal{M}_k} r_m$.

\textbf{Parallel processing on different modalities.} 
Multi-modal models often have multiple encoders depending on the modalities that are used in input data. Although they are one set of input data, they are processed independently in each different encoder. Our split architecture enables encoders to be processed in parallel due to its modular structure. On the other hand, a head module is to process task-specific requests, and thereby each task has only one for each task, which can only be processed after all encodings are completed.

\vspace{-0.5mm}
\subsection{Shareable Architecture in Multi-Task Inference}
\vspace{-0.5mm}
\label{sec:split-and-share}

\begin{figure}[!t]
    \centering
    \includegraphics[width=0.8\columnwidth]{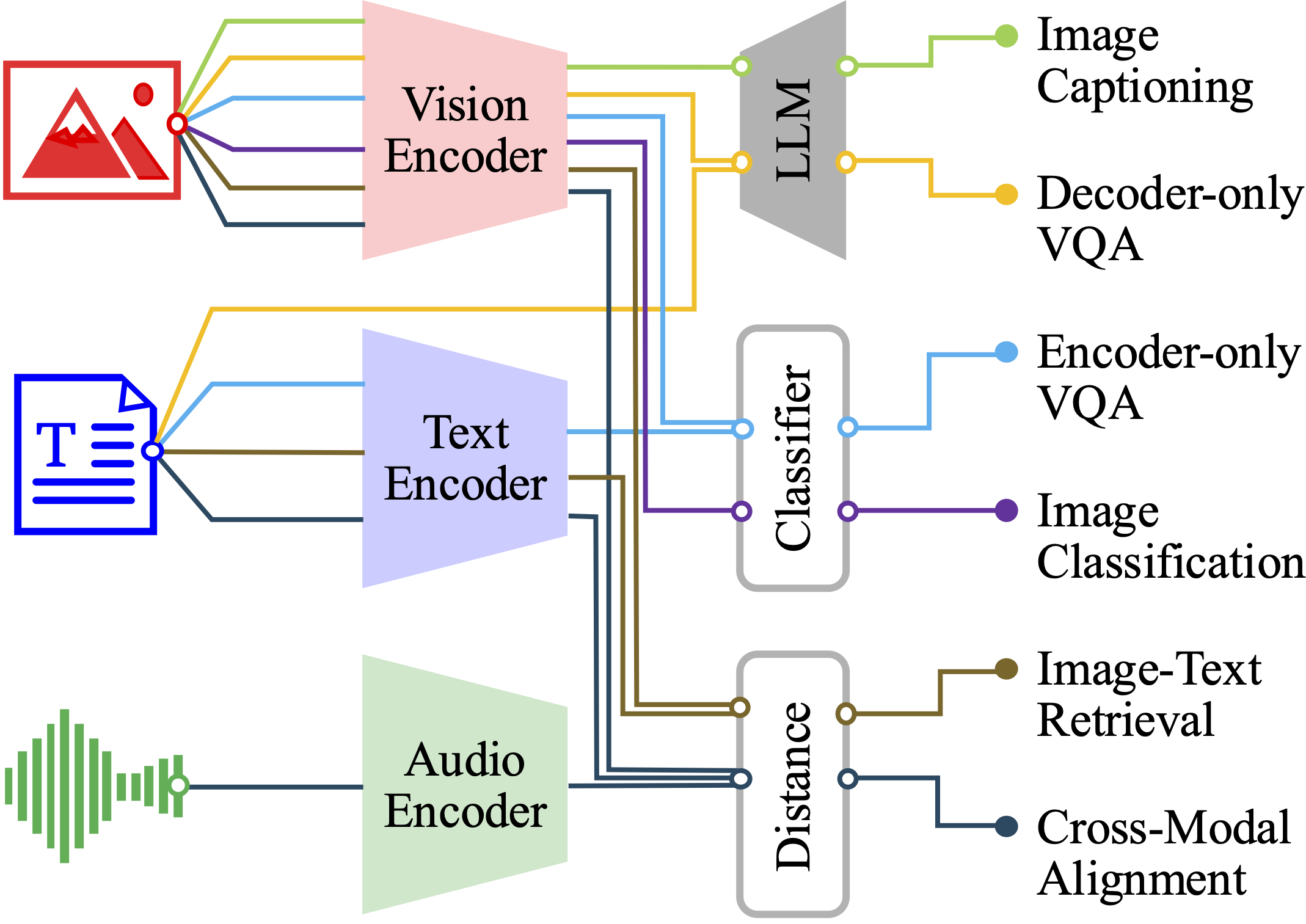}
    \vspace{-0.5mm}
    \caption{Overall architecture of \emph{\ours} to \emph{split} into functional modules and \emph{share} across diverse tasks.}
    \vspace{-5mm}
    \label{fig:overview}
\end{figure}

Given the increasing demands for diverse tasks, from uni-modal to multi-modal, it often requires a dedicated model for each task, incurring excessive deployment costs; proportionally increasing to the number of tasks. However, as illustrated in \reffig{vl-task-arch}, multi-modal tasks often have similar functions. 

\textbf{Module sharing.} 
Inspired by Insights \ref{insight:interchange} and \ref{insight:sharable}, we reuse the already deployed functional modules for the common modules and only allocate additional resources for uncommon modules, as shown in \reffig{overview}. It has extendibility in various multi-modal tasks, where the number of encoders and task heads each corresponds only up to the number of modalities and the number of tasks. By reusing existing modules and loading additional ones as needed for specific tasks, \emph{split-and-share} architecture can significantly reduce deployment costs for edge devices, particularly as the number of tasks increases.

Considering module sharing in the multi-task setting, we can compute the total deployment cost. 
Let $\mathcal{K}$ be the set of all the inference models we consider. The entire module set for all the models to be deployed is $\mathcal{M} := \bigcup_{k \in \mathcal{K}} \mathcal{M}_{k}$. With our module sharing in multi-task inference, the worst-case total deployment cost is $O(c \cdot r)$, where $c$ is the number of distinct modules. In contrast, the cost without sharing is $O(|\mathcal{M}| \cdot r)$ with duplicate module deployment, where $c << |\mathcal{M}|$.

%% file: 5_placement_v4.tex
By sharing common modules across tasks, we can significantly reduce the total memory requirement. However, this shared module approach introduces a new trade-off between memory usage and latency. While memory usage decreases as more modules are shared, a bottleneck arises when multiple requests access the same module simultaneously, increasing the latency. A straightforward way to avoid the bottleneck is to deploy each module for each task independently without sharing. This approach, however, would waste memory resources when requests for certain tasks are infrequent. Therefore, the resource allocation problem faces unique challenges compared to a single task setting. To solve this, we first formulate the placement and request routing problem in \refsec{formulation} and provide a module-level greedy solution in \refsec{solution}.

\vspace{-0.5mm}
\subsection{Problem Formulation}\label{sec:formulation}
\vspace{-0.5mm}

Let $\mathcal{N}$ be the set of devices, where each device $n \in \mathcal{N}$ has a memory capacity $R_n$. 
We aim to place the entire module set $\mathcal{M}$ from all inference models in $\mathcal{K}$ over the devices in $\mathcal{N}$. 
Let $\mathcal{Q}$ denote a set of requests that arrive sequentially. For each request $q \in \mathcal{Q}$, $k(q) \in \mathcal{K}$ be its required model; accordingly, let $\mathcal{M}^{\mathrm{enc}}_{k(q)}$ be the required encoders, $h_{k(q)}$ be the task head, and $\mathcal{M}_{k(q)}$ be the whole module set. Let $n_q$ be the source device that initiates request $q$, and multiple requests can be made for the same model. 
Let $x_{m, n} \in \{0, 1\}$ denote the binary placement decision variable, indicating whether module $m$ is placed on device $n$ or not, and let $\mathbf{x}$ denote the placement decision variable matrix, i.e., $\mathbf{x} := (x_{m, n})_{m \in \mathcal{M}, n \in \mathcal{N}}$. Similarly, let $y^q_{m, n} \in \{0, 1\}$ denote the binary routing decision variable for request $q$, indicating whether to route the request to module $m$ on device $n$.
Let $\mathbf{y}^q$ denote the routing decision variable for request $q$, i.e., $\mathbf{y}^q := (y^q_{m, n})_{m \in \mathcal{M}, n \in \mathcal{N}}$; let $\mathbf{y}$ denote the routing decision variable matrix, i.e., $\mathbf{y} := (\mathbf{y}^q)_{q \in \mathcal{Q}}$.
We use $a_{m, n}$ to represent the capacity of device $n$ in processing sub-requests for module $m$ if it hosts the module (e.g., the computation capability or the batch size).

Requests arrive at \emph{model} level, while the placement is performed at \emph{module} level. Each inference request demands a model consisting of multiple inter-dependent modules that need to be traversed in a certain order. Here, we aim to optimize the end-to-end latency, defined as $t^{\mathrm{total}}(\mathbf{y}^q)$ in Eq.~\eqref{eq:MPRR-R-total}:
\begin{align}
t^{\mathrm{total}}(\mathbf{y}^q) & := t^{\mathrm{enc}} (\mathbf{y}^q) + t^{\mathrm{head}} (\mathbf{y}^q), \label{eq:MPRR-R-total} \\
t^{\mathrm{enc}} (\mathbf{y}^q) & := \max_{m \in \mathcal{M}^{\mathrm{enc}}_{k(q)}} \Big\{ \sum\limits_{n \in \mathcal{N}} y^q_{m, n} \Big(t^{\mathrm{comm}}_{m, n_q, n} + t^{\mathrm{comp}}_{m, n} \notag \\
& \hspace{21mm} + \sum_{n' \in \mathcal{N}} y^q_{h_{k(q)}, n'} \cdot t^{\mathrm{comm}}_{h_{k(q)}, n, n'} \Big) \Big\}, \label{eq:MPRR-R-enc} \\
t^{\mathrm{head}} (\mathbf{y}^q) & := \sum\limits_{n' \in \mathcal{N}} y^q_{h_{k(q)}, n'} \cdot t^{\mathrm{comp}}_{h_{k(q)}, n'}. \label{eq:MPRR-R-head}
\end{align}
Specifically, the end-to-end latency becomes the sum of two terms: encoder latency $t^{\mathrm{enc}}(\mathbf{y}^q)$ in Eq.~\eqref{eq:MPRR-R-enc}; and task head latency $t^{\mathrm{head}}(\mathbf{y}^q)$ in Eq.~\eqref{eq:MPRR-R-head}. Here, the encoder latency $t^{\mathrm{enc}}(\mathbf{y}^q)$ consists of three main parts: (i) the user data transmission latency of sending the input data for module $m$ from the source node $n_q$ to encoder device $n$, denoted by $t^{\mathrm{comm}}_{m, n_q, n}$; (ii) the computation time of encoding by module $m$ on device $n$, denoted by $t^{\mathrm{comp}}_{m, n}$; and (iii) the output transmission latency of sending the encoding output from device $n$ hosting the encoder module to some device $n'$ hosting the required task head $h_{k(q)}$, denoted by $t^{\mathrm{comm}}_{h_{k(q)}, n, n'}$. Due to parallel processing, the encoding latency for request $q$ takes the maximum over all the required encoder modules in $\mathcal{M}^{\mathrm{enc}}_{k(q)}$, i.e., it is determined by the slowest encoder. Once all the encoding outputs are received, the task head latency is defined by the computation time for head $h_{k(q)}$ on the device $n'$ hosting the head, denoted by $t^{\mathrm{comp}}_{h_{k(q)}, n'}$.

We aim to find the best routing strategy to minimize total (hence the average) latency over all requests in $\mathcal{Q}$ as follows:
\begin{subequations}\label{eq:MPRR}
\vspace{-4mm}
\begin{align}
    \min\limits_{\mathbf{x}, \mathbf{y}
    } & \sum\limits_{q \in \mathcal{Q}} t^{\mathrm{total}} (\mathbf{y}^q) \label{eq:MPRR-obj} \\
    \mbox{s.t.}~& \sum\limits_{q \in \mathcal{Q}} y^{q}_{m, n} \le a_{m, n} \cdot x_{m, n}, &\hspace{-7mm}  \forall m \in \mathcal{M}, \forall n \in \mathcal{N}, \label{eq:MPRR-placed}\\
    & \sum\limits_{n \in \mathcal{N}} y^{q}_{m, n} = 1, &\hspace{-7mm} \forall q \in \mathcal{Q}, \forall m \in \mathcal{M}_{k(q)}, \label{eq:MPRR-routingonce} \\
    & \sum_{m \in \mathcal{M}} x_{m, n} r_{m} \leq R_{n}, &\hspace{-7mm} \forall n \in \mathcal{N}, \label{eq:MPRR-memory} \\
    & x_{m, n}, y^{q}_{m, n} \in \{0, 1\}, &\hspace{-7mm} \forall q \in \mathcal{Q}, \forall m \in \mathcal{M}, \forall n \in \mathcal{N}. \label{eq:MPRR-decision}
\vspace{-4mm}
\end{align}
\end{subequations}
Constraint~\eqref{eq:MPRR-placed} ensures that a request can only be routed to a module that is placed, i.e., $x_{m, n}=1$, and the total requests routed to module $m$ on device $n$ cannot exceed capacity $a_{m, n}$. 
Constraint~\eqref{eq:MPRR-routingonce} enforces that each request is routed to required modules only once. 
Different from traditional split models, where submodels are processed sequentially, and the total latency is the simple summation of all computation and communication latencies, we allow for parallel processing across different modalities (thus the maximum operation in \eqref{eq:MPRR-R-enc}). 
Constraint~\eqref{eq:MPRR-memory} ensures that the memory capacity of device $n$ is not exceeded.
In Problem~\eqref{eq:MPRR}, $\mathbf{x}$ and $\mathbf{y}$ are primary decision variables; $a_{m, n}$, $r_m$, and $R_n$ are all given constants.

This problem introduces two new challenges: \emph{First}, it involves requests arriving at the model level, while placement decisions are made at the module level. Thus, a module can be shared across different models, introducing an additional cross-model dependency. \emph{Second}, while some existing works~\cite{huang2024distmm,zhang2024disttrain,feng2024optimus} also consider parallel processing, they primarily focus on pipelining multiple batches. In contrast, we propose to perform parallel processing over different modalities within the same request.

\vspace{-1mm}
\subsection{Proposed Solution}\label{sec:solution}
\vspace{-0.5mm}

Given the high placement costs and memory loading time to download the modules onto devices, migrating or replicating the modules may incur a significant overhead compared to the actual inference time.\footnote{The model download time of CLIP ViT-B/16 from HuggingFace~\cite{huggingface} is 9.36~sec (depending on network stability), and the model loading time on Tesla P40 GPU is 11.08~sec, totaling 20.44~sec for placing once. A single inference request on the Tesla P40 GPU takes 2.44~sec.} 
Therefore, we solve the problem sequentially: the module placement in a larger time scale; and the per-request inference routing in a smaller time scale. 

\textbf{Greedy module placement.}
To determine the placement strategy aiming to reduce the end-to-end latency as in Problem~\eqref{eq:MPRR}, we adopt a greedy heuristic due to its simplicity yet efficiency in multi-modal architectures with multiple distinct modules. These functional modules have unique memory requirements and widely varying computation times, showing significant differences in latency, especially for compute-intensive modules.\footnote{Processing a text encoder in CLIP ViT-B/16 takes about 3~sec on a laptop but 43~seconds on a Jetson Nano device.} As the module size increases, the inference time gap tends to become larger, making it essential to handle compute-intensive modules effectively to minimize overall latency. Given that communication latency is minimal compared to the computation time in our edge network scenarios (see~\reffig{detailed-latency}), our greedy approach focuses on computation time, which dominates the end-to-end latency. To accelerate inference, we first prioritize to place the module that requires larger memory, i.e., $\max_{m \in \mathcal{M}} r_m$.

To determine a device-to-module placement decision $x_{m, n}$, we allocate the modality encoding and the task head modules one by one in a greedy manner. 
In placing encoders, our greedy approach first deploys the module $m \in \mathcal{M}^{\mathrm{enc}}$ on the device $n$ with the shortest completion time. The completion time $t^{\mathrm{place}}_{m, n}$ is the accumulated computation time of the module $m$ and the already deployed modules $m'$ on device $n$, defined as:
\begin{align}\label{eq:accum-time-enc}
    t^{\mathrm{place}}_{m, n} := t^{\mathrm{comp}}_{m, n} +\hspace{-1mm}\sum_{m' \in \mathcal{M}} x_{m', n} \cdot t^{\mathrm{comp}}_{m', n}, &\hspace{1mm} \forall m \in \mathcal{M}^{\mathrm{enc}},
\end{align}
where $x_{m', n}$ is 1 if module $m'$ is deployed on devices $n$. If the resource of device $n^* = \argmin_{n\in N} t^{\mathrm{place}}_{m,n}$ is enough, i.e., $r_{m} \leq R_{n^*}$, we deploy the module $m$ on device $n^*$, i.e., $x_{m, n^*}=1$. If the device with the shortest completion time cannot load the module due to resource limits, it searches for the next device with the next shortest completion time that has enough resources.

On the other hand, task head module $m \in \mathcal{M}^{\mathrm{head}}$ can be processed only after all encoders are processed. We do not consider the accumulated time but only the task head computation time. We thus prioritize the device with the smallest computation time as follows:
\begin{align}\label{eq:accum-time-head}
    t^{\mathrm{place}}_{m, n} := t^{\mathrm{comp}}_{m, n}, & \hspace{5mm} \forall m \in \mathcal{M}^{\mathrm{head}}.
\end{align}
By ensuring these modules are allocated to the most powerful devices with minimal computation completion time while maintaining parallelism, our greedy approach ensures that the worst-case module processing time is minimized.

We do not explicitly include further partitioning with the modules in our placement, but if the module cannot be loaded on any devices, we can further apply compression or DNN/LLM partitioning techniques to make the modules more lightweight. After leveraging such techniques, we can search the devices for partitioned modules (as one module) using our greedy placement approach. If we have remaining resources, we replicate the modules with larger memory requirements.

\begin{algorithm}[t]
\footnotesize
\caption{Greedy Module Placement and Routing in \emph{\ours}} \label{alg:our-algo}
\begin{algorithmic}[1]
\floatname{algorithm}{Procedure}

\renewcommand{\algorithmicrequire}{\textbf{Input:}}
\STATE \algorithmicrequire~ $\mathcal{N}$, $\mathcal{M}$, $\mathbf{R}$, $\mathbf{r}$, $\mathbf{t}^{\mathrm{comp}}$\\
\vspace{1mm}

\STATE Initialize $x_{m, n}=0$; \hfill$\triangleright$ Greedy module placement
\STATE {\bf for} module $m \in \mathcal{M}$ {\bf do} \hspace{3mm}// In descending order of memory requirements \\
    \STATE\hspace{1\algorithmicindent}{Sort $\mathcal{N}$ by the completion time $t^{\mathrm{place}}_{m, n}$ in Eq.~\eqref{eq:accum-time-enc} and Eq.~\eqref{eq:accum-time-head};}
    \STATE\hspace{1\algorithmicindent}{\bf for} device $n \in \mathcal{N}$ {\bf do} \hspace{2.5mm}// In ascending order of completion time \\
        \STATE\hspace{2\algorithmicindent}{\bf if} 
        $r_m \leq R_n$ {\bf then}\hspace{4mm}// If having enough memory,\\
            \STATE\hspace{3\algorithmicindent}{$x_{m, n}=1$; \hspace{6.5mm} // Place module $m$ on device $n$} 
            \STATE\hspace{3\algorithmicindent}$R_n = R_n - r_m$; \hspace{0mm}// Decrease the available memory
            \STATE\hspace{3\algorithmicindent}break;
        \STATE\hspace{2\algorithmicindent}{\bf end if}
    \STATE\hspace{1\algorithmicindent}{\bf end for}
\STATE {\bf end for}
\vspace{1mm}

\STATE Load modules on each device; \hfill$\triangleright$ Per-request inference routing
\STATE{\bf while} $q=$ has\_next$()$ {\bf do} \hspace{11mm} // For request $q$ from task $k$
    
    \STATE\hspace{1\algorithmicindent}{\bf for} $m \in \mathcal{M}^{\mathrm{enc}}_{k(q)}$ in parallel {\bf do} \hspace{2mm}// Parallel processing on encoders
        \STATE\hspace{2\algorithmicindent}Process encoding on the device of $\min_{n \in \mathcal{N}_m} t^{\mathrm{comp}}_{m, n}$ in Eq.~\eqref{eq:time-routing};
    \STATE\hspace{1\algorithmicindent}{\bf end for}
    \STATE\hspace{1\algorithmicindent}Process task head on the device of $\min_{n \in \mathcal{N}_m} t^{\mathrm{comp}}_{h_{k(q)}, n}$ in Eq.~\eqref{eq:time-routing};

    \hspace{1\algorithmicindent}\hspace{39mm}// Processing on task head
\STATE{\bf end while}

\end{algorithmic}
\end{algorithm}

\textbf{Per-request parallel routing.}
Based on the placement decision, we load all modules on each device and then process routing for each inference request. Here, we consider parallel routing over multiple encoders within a single request.

For each inference request $q$ from model $k$, we send each input modality data (only if the requester device and the device to encode the data are different). We first select the route for the encoder module set $\mathcal{M}^{\mathrm{enc}}_{k(q)}$ and then for the task head module $h_{k(q)}$. Similar to placement, we select the device $n$ with the shortest computation time for module $m$ as follows:
\begin{align}\label{eq:time-routing}
    \argmin_{n \in \mathcal{N}_m} t^{\mathrm{comp}}_{m, n}, & \hspace{5mm} \forall m \in \mathcal{M}_{k(q)},
\end{align}
where $n \in \mathcal{N}_m$ is the device having the module $m$ (i.e., $x_{m, n} = 1$). 
Once routing is determined, we send the data with a modality that takes longer in the encoding first to initiate the longest encoding as early as possible.  All devices in charge of any encoder process the encoding for each modality in parallel. After all encodings are completed, the results are sent to the device in charge of the task head module. 

Furthermore, to reduce the overall latency on a series of multiple requests, we process requests one-by-one but can initiate the next request as soon as encoders are available, similar to pipelining~\cite{narayanan2019pipedream}.
The algorithm is detailed in \refalg{our-algo}.

%% file: 6_experiments_v3.tex
We have validated \emph{\ours} on 14 public models across five multi-modal tasks with 10 benchmarks. 
We used \texttt{socket} programming for transmitting input data and embeddings among devices.
In experiments, we aim to validate the following research questions in terms of the memory efficiency, latency, and accuracy of \emph{\ours} for multi-modal models:
\begin{itemize}[leftmargin=8mm]
    \item[\textit{Q1:}] How efficiently does our split architecture reduce the memory requirements on multi-modal inference? 
    \item[\textit{Q2:}] Can \emph{\ours} provide reasonable (or even better than cloud) latency only using edge devices?
    \item[\textit{Q3:}] Is there any accuracy drop to make multi-modal models runnable on edge devices?
    \item[\textit{Q4:}] How efficiently does our shared architecture save memory usage across multi-task inference?
\end{itemize}


\begin{table}[!t]
    \centering
    \caption{Device specification. Four resource-constrained edge devices are deployed in a home network, and a server with a GPU is deployed out of a home network. 
    }
    \label{tab:device-spec}
    \begin{tabular}{llllllll}
        \toprule
         & CPU (RAM) & GPU (VRAM) \\\midrule
        Server & Intel Xeon Gold 5115 (33.7~GB) & Tesla P40 (23.9~GB) \\\midrule
        Desktop & Intel i7-13700 (31.7~GB) & -- \\
        Laptop & Apple M3 Pro (18.0~GB) & -- \\
        Jetson A & ARMv8 Processor (4.1~GB) & -- \\
        Jetson B & ARMv8 Processor (4.1~GB) & -- \\
        \bottomrule
    \end{tabular}
    \vspace{-1mm}
\end{table}

\textbf{Device and network settings.}
As described in \reftab{device-spec}, we used four resource-constrained devices deployed in a home network in the Personal Area Network (PAN)-level, including a desktop, a laptop, and two 4GB Jetson Nano (P-3450) devices~\cite{nvidia_jetson_nano}. A server is located out of PAN at the level of the Metropolitan Area Network (MAN). The desktop is connected in wired, whereas the laptop is connected in wireless as general use. Laptop and Jetson A are connected wirelessly using Wi-Fi (IEEE 802.11). We did not control or restrict any network traffic to reflect realistic daily network conditions. To implement cloud computing, we used one server equipped with a GPU. It should be noted that different from typical cloud servers, which have numerous users, we used a dedicated server. Thereby, while ChatGPT or Gemini servers take around 13-15~ms for each packet, our server only takes 4-5~ms on average. We set Jetson A (wireless Jetson Nano) as the default requester, which has input data and initiates inference.


\begin{table}[!t]
    \centering
    \caption{Functional modules in multi-modal tasks, where `$||$' denotes that parallel processing is available.}
    \label{tab:task-arch}
    \setlength{\tabcolsep}{2.2pt}
    \begin{tabular}{lllllll}
        \toprule
        & \multicolumn{3}{c}{Encoder} & \multicolumn{3}{c}{Task Head} \\\cmidrule(lr){2-4}\cmidrule(l){5-7}
        & Vision & Text & Audio & LLM & Distance & Classifier \\ \midrule
        Image-Text Retrieval ($||$) & \multicolumn{1}{c}{\checkmark} & \multicolumn{1}{c}{\checkmark} & & & \multicolumn{1}{c}{\checkmark} & \\
        Encoder-Only VQA ($||$) & \multicolumn{1}{c}{\checkmark} & \multicolumn{1}{c}{\checkmark} &  &  &  & \multicolumn{1}{c}{\checkmark} \\
        Decoder-only VQA & \multicolumn{1}{c}{\checkmark} & & & \multicolumn{1}{c}{\checkmark} \\
        Cross-Modal Alignment ($||$) & \multicolumn{1}{c}{\checkmark} & \multicolumn{1}{c}{\checkmark} & \multicolumn{1}{c}{\checkmark} &  & \multicolumn{1}{c}{\checkmark} \\
        Image Classification & \multicolumn{1}{c}{\checkmark} & & & & & \multicolumn{1}{c}{\checkmark}\\
        \bottomrule
    \end{tabular}
\end{table}

\begin{table}[!t]
    \centering
    \caption{Model size for each functional module. For example, flint-v0.5-1B FM for VQA task uses ViT-L/16@336px and TinyLlama-1.1B, consisting of 304M+1.1B=1.4B parameters. Please refer to the architecture information in \reftab{fm-arch}.}
    \label{tab:runnable-module}
    
    \begin{tabular}{lll}
        \toprule
        {Functional Module}
        & {Module}
        & \# Param \\ \midrule
        \multirow{10}{*}{Vision Encoder}
        & ResNet-50 & 38M  \\
        & ResNet-101 & 56M \\
        & ResNet-50x4 & 87M \\
        & ResNet-50x16 & 168M \\
        & ResNet-50x64 & 421M \\
        & ViT-B/32 & 88M \\ 
        & ViT-B/16 & 86M \\
        & ViT-L/14 & 304M \\ 
        & ViT-L/14@336 & 304M \\ 
        & OpenCLIP ViT-H/14 & 630M \\ \midrule
         \multirow{2}{*}{Text Encoder} & CLIP TRF & 38-85M \\ 
        & OpenCLIP TRF & 302M \\ \midrule
        Audio Encoder & ViT-B & 85M \\ \midrule
        \multirow{3}{*}{Language Model} & Vicuna-7B & 7B \\
        & Phi-3-Mini & 3.8B \\
        & TinyLlama-1.1B & 1.1B \\
        \bottomrule
    \end{tabular}
    \vspace{-2mm}
\end{table}

\textbf{Tasks and benchmarks.}
We implemented five different multi-modal tasks with ten benchmarks: 1) image-text retrieval task using Food-101~\cite{bossard2014food}, CIFAR-10~\cite{cifar10}, CIFAR-100~\cite{cifar10}, Country-211~\cite{thomee2016yfcc100m}, and Flowers-102~\cite{nilsback2008automated}; 2) encoder-only VQA using MS COCO~\cite{lin2014microsoft}; 3) decoder-only VQA using VQA-v2~\cite{goyal2017making}, ScienceQA~\cite{lu2022learn}, and TextVQA~\cite{singh2019towards}; 4) cross-modal alignment using As-A~\cite{gemmeke2017audio}; and 5) image classification using Food-101~\cite{nilsback2008automated}. As shown in \reftab{task-arch}, some tasks such as image-text retrieval, encoder-only VQA, and cross-modal alignment have multiple encoders, implying a parallel processing is available. In contrast, some tasks, such as decoder-only VQA or image classification, have only one encoder, which means they cannot benefit from parallel processing. We used CLIP ViT-B/16 for the image-text retrieval task as the default unless otherwise noted. The model size for each modality-wise encoder and language model is in \reftab{runnable-module}.

\textbf{Metrics.}
We use three metrics: 1) accuracy: to maintain the zero-shot inference performance using pretrained models without any modification or additional fine-tuning; 2) latency: to reduce the inference time; and 3) the number of parameters: to reduce the model size deployed on a single device. Here, we applied two different latency measures: 1) inference latency: to evaluate the pure inference time from sending input data to corresponding devices to generating the output; and 2) end-to-end latency: to show the entire time including the time to load the model on devices--which may vary depending on the device hardware. As latency is affected by the instantaneous network status, we averaged over five trials.

\begin{table}[!t]
    \centering
    \caption{Deployment cost and latency using various multi-modal tasks and architectures, where Cloud and Local mean the Centralized inference only on the cloud and the Jetson.}
    \label{tab:diff-architecture}
    \setlength{\tabcolsep}{3.2pt}
    \begin{tabular}{lllllllll}
        \toprule
        Architecture & \multicolumn{3}{c}{\# Param} & \multicolumn{3}{c}{Inference Time (sec)} \\ \cmidrule(lr){2-4}\cmidrule(l){5-7}
         & Centralized & \multicolumn{2}{c}{\emph{\ours}} & Cloud & Local & \emph{\ours}\\ \midrule
        \multicolumn{6}{l}{(Image-Text Retrieval)} \\
        CLIP ResNet-50 & 76M & \textbf{38M} & (-50\%) & 2.73 &53.23 & \textbf{2.32}\\  
         CLIP ResNet-101 & 94M & \textbf{56M} & (-40\%) &2.63	 & 48.87	& \textbf{2.39}	\\  
         CLIP ResNet-50x4 & 146M & \textbf{87M} & (-40\%) &2.64 &64.54	 &3.07	 \\  
         CLIP ResNet-50x16 & 253M	& \textbf{168M} & (-34\%) &2.65 & -- &4.56 \\  
         CLIP ResNet-50x64	& 572M & \textbf{421M} & (-26\%) &2.92	 &-- & 6.50 \\  
         CLIP ViT-B/32 & 126M& \textbf{88M} & (-30\%) & 2.42	 & 44.26 & 2.49	 \\  
         CLIP ViT-B/16 & 124M & \textbf{86M} & (-31\%) & 2.44  & 45.19 & 2.48	\\  
         CLIP ViT-L/14 & 389M & \textbf{304M} & (-22\%) & 2.61 & -- & 4.46 \\  
         CLIP ViT-L/14@336 & 389M & \textbf{304M} & (-22\%) &2.65	 &-- &4.51\\  
         \midrule
         
        \multicolumn{6}{l}{(VQA)} \\
        Encoder-only (Small) & 124M & \textbf{86M} & (-31\%) &1.23	 & 6.28 & \textbf{0.50}\\
         Encoder-only (Large) & 389M & \textbf{304M} & (-31\%) &1.50 & -- &\textbf{1.23}	 \\ 
         \midrule 
         
        \multicolumn{6}{l}{(Cross-Modal Alignment)} \\
        ImageBind & 1.0B & \textbf{630M} & (-37\%) &2.44	 &-- & \textbf{2.34}\\ 
         \bottomrule
    \end{tabular}
\end{table}

\textbf{Baselines.} 
We compared with the following baselines: 1) Optimus~\cite{feng2024optimus}; 2) DistMM~\cite{huang2024distmm}: multi-modal training approaches by extracting its partitioning strategy \emph{in a few tasks}--Optimus is designed only for VQA, while DistMM only considers image-text retrieval tasks; and 3) Megatron-LM~\cite{shoeybi2019megatron}: model parallelism approach applied to each module. We also provided several strong baselines: 4) \emph{Centralized} server and local: loading all modules on a single powerful GPU-equipped server or a local device; and 5) \emph{Upper}: an optimal placement solution with the least latency in a brute-force manner.
There are no existing baselines that address the complexities of multi-modal, multi-task inference. Existing intra-module partitioning, compression, or multi-modal training methods are not direct competitors but can be applied orthogonally to our inter-module partitioning (please refer to \refsec{related}).

\begin{table}[!t]
    \centering
    \caption{Comparison of deployment cost and latency.}
    \label{tab:runnable-latency}
    \setlength{\tabcolsep}{5pt}
    \begin{tabular}{lllll}
        \toprule
         \multicolumn{2}{l}{Deployment} & \# Param & \multicolumn{2}{c}{Latency (sec)} \\ \cmidrule{4-5}
          & & & Inference & End-to-End \\ \midrule
         \multirow{5}{*}{Centralized} & Server & \multirow{5}{*}{124M} & 2.44 &13.53 \\
          & Server (w/o GPU) &  & 6.70& 17.78  \\
          & Desktop & &3.46 & 4.95	  \\
          & Laptop & &3.02 & 5.31 	 \\
          & Jetson & & 45.19 &60.37 \\\midrule
         \emph{\ours}& & \multirow{2}{*}{\textbf{86M}}  & \textbf{2.48} & \textbf{4.76} \\
         \multicolumn{2}{l}{\emph{\ours} (w/o Parallel Processing)} & & 3.03 & 5.32 \\
         \bottomrule
    \end{tabular}
    \vspace{-3mm}
\end{table}

\subsection{Split Architecture in Multi-Modal Inference}\label{sec:eval-split}

We first verified the memory and latency efficiency of our \emph{split} architecture within a single multi-modal model.

\textbf{Q1: Saved resource via distributed deployment.}
We showed how our split architecture contributes to resource-saving for on-device AI. In \reftab{diff-architecture}, the resource required in various architectures in centralized deployment and distributed deployment in \emph{\ours}. For example, in the most effective case among all models, an image-text retrieval task with CLIP ResNet-50 requires 76.2M parameters to deploy in a centralized manner. However, as we deploy each module on each device, the memory necessary for a single device is reduced by 49.6\%. Compared to optimal placement (\emph{Upper}), our module-level greedy placement method achieves the optimal latency in 89 cases among 95 cases (19 combinations of benchmarks and models $\times$ 5 trials), resulting in 93.7\%.

In particular, in CLIP ViT-B/16 case in \reftab{runnable-latency}, the model consisting of 124M parameters is reduced to 86.2M and 37.8M. It enables each device to use fewer resources allocated for the task, giving more room for other jobs. This distributed deployment also led to models runnable on edge devices that were not able to be processed. From \reftab{diff-architecture}, Jetson Nano cannot run the entire models for some architectures such as CLIP ResNet-50x16 or ImageBind (denoted as `--'), but distributed modules can be loaded across different devices. 

\begin{implication}[Memory efficiency I]
    Our split architecture not only enables large models runnable on edge devices but also reduces deployment cost by up to 49.6\%.
\end{implication}


\begin{figure}[!t]
    \centering
    \includegraphics[width=0.99\linewidth]{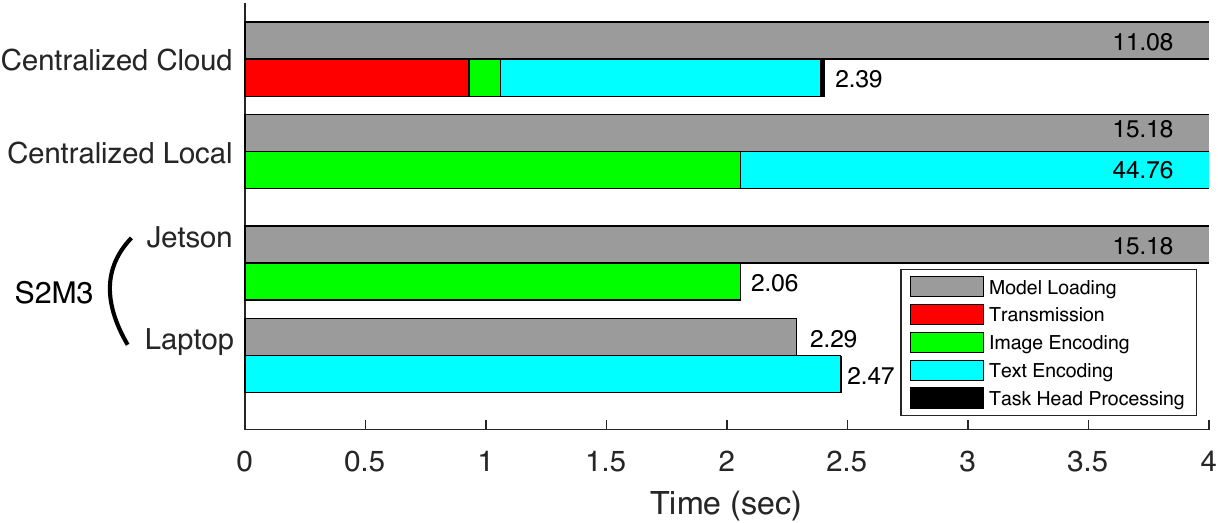}
    \caption{Inference timeline using CLIP ViT-B/16 for image-text alignment, where Jetson and a laptop have a vision and a text encoder, respectively. 1) Jetson (requester) sends an input prompt to Laptop; 2) Jetson and Laptop process the encoding for each modality in parallel; 3) Laptop sends encoded text features back to Jetson; and 4) finally, Jetson processes a task head. Here, the transmission and the task head processing take a minimal time, nearly invisible in this case.}
    \vspace{-5mm}
    \label{fig:detailed-latency}
\end{figure}

\textbf{Q2: Reduced latency via parallel processing.}
In terms of inference latency, as shown in \reftab{diff-architecture}, the inference time is reduced by up to 56.91\% in Encoder-only (Small) for the VQA task, compared to the centralized server. Of course, depending on the compute-intensiveness of encoders and task heads, there are some cases, such as ResNet-50x64 for image-text retrieval task, which needs more inference time compared to the centralized cloud. ResNet-50x64 has a very large vision encoder of 421M and a relatively small text encoder of 151M, and thereby the overall inference time is dominated by the vision encoding time, and thus the effect of parallel processing is relatively small. Nevertheless, we can make it runnable on edge devices that was originally not available.

More in detail, for CLIP ViT-B/16 in \reftab{runnable-latency}, a requester (Jetson) takes too long to do one-shot inference with 45.19s. By loading the compute-intensive module to a more powerful device based on our greedy deployment, \emph{\ours} highly reduces the inference time from 45.19s to 2.48s, comparable to 2.44s in a centralized cloud. Even with available neighboring resources, if we just do the inference in a centralized manner, e.g., let the desktop do the entire inference, it cannot benefit from parallel processing (unless installing more processors in a single device), implying that any centralized approaches are not desirable. Taking a closer look at the detailed inference timeline in \reffig{detailed-latency} (shown using only the Jetson and Laptop for visual clarity, while all other results use five devices as default setting), as image and text encodings are run simultaneously on different devices of Jetson and Laptop and accelerate the inference, ideally proportionally to the number of modalities faster, making comparable to the cloud.

\begin{implication}[Reduced Latency]
    We can reduce the inference time by up to 56.9\%, ideally proportionally to the number of modalities, by processing the multiple modalities in parallel.
\end{implication}


\begin{table}[!t]
    \centering
    \caption{Accuracy on \emph{\ours} across various benchmarks.}
    \vspace{-1mm}
    \label{tab:diff-benchmark}
    \begin{tabular}{llllllllll}
        \toprule
        Architecture (Task) & Benchmark & \multicolumn{2}{c}{Accuracy (\%)} \\\cmidrule(lr){3-4}
          & & \emph{\ours} & Reported \\ 
        \midrule
        
         \multirow{5}{*}{\shortstack[l]{CLIP ViT-B/16\\(Image-Text Retrieval)}} & Food-101 & 87.7 & 89.2 \\ 
          & CIFAR-10 & 90.8 & 91.6 \\ 
         & CIFAR-100 & 66.9 & 68.7 \\ 
         & Country-211 & 22.4 & 23.3 \\ 
         & Flowers-102 & 71.0 & 70.4 \\\midrule
         
         \multirow{5}{*}{\shortstack[l]{CLIP ViT-L/14@336\\(Image-Text Retrieval)}} & Foods-101 & 93.2 & 93.8 \\ 
        & CIFAR-10 & 94.9 & 95.7 \\ 
         & CIFAR-100 & 74.3 & 77.5 \\ 
         & Country-211 & 33.9 & 34.9 \\ 
         & Flowers-102 & 77.1 & 78.3\\ \midrule
        
        \multirow{3}{*}{\shortstack[l]{Flint-v0.5-1B\\(VQA)}} & VQA-v2 & 70.2 & -- \\ 
          & ScienceQA & 41.2 & --\\ 
         & TextVQA & 35.6 & --\\\midrule
         \multirow{3}{*}{\shortstack[l]{LLaVA-v1.5-7B\\(VQA)}} & VQA-v2 & 78.1 & 78.5\\ 
          & ScienceQA & 69.4 & 70.4\\ 
         & TextVQA & 57.3 & --\\ 
        \bottomrule
    \end{tabular}
    \vspace{-4mm}
\end{table}

\textbf{Q3: Unimpacted accuracy of pretrained models.}
We leveraged the models without modification and validated the zero-shot inference accuracy on the same model. As shown in \reftab{diff-benchmark}, \emph{\ours} does not sacrifice accuracy while making models available on edge devices. Different from approaches that modify the model by customizing into the target domain, where the accuracy is affected via modification, we are using the same architecture, thereby showing very similar accuracy (ideally should be the same, but a marginal accuracy drop appears in some cases, which is not caused by our architecture but by runtime variability.) with the reported accuracy. 

\begin{implication}[Maintained accuracy]
    We can reduce the resource usage and latency while not harming the accuracy of the original models.
\end{implication}


\begin{table}[!t]
    \centering
    \caption{Device availability, where S, D, L, J-B, and J-A represent server, desktop, laptop, Jetson B (wired), and Jetson A (wireless), respectively.}
    \label{tab:device-availability}
    \setlength{\tabcolsep}{4pt}
    \begin{tabular}{lcccccll}
        \toprule
         & S & D & L & J-B & Requester: J-A & Latency (sec) & \# Param \\ \midrule
        \multirow{2}{*}{Centralized} & \checkmark & & & & \checkmark & 2.44 & \multirow{2}{*}{124M}\\ 
        & & & & & \checkmark & 45.19 & \\ \midrule
        \multirow{3}{*}{\emph{\ours}} & & & & \checkmark & \checkmark &\textbf{42.70} & \\
         & & & \checkmark & \checkmark & \checkmark & \textbf{2.49}& \textbf{86M} \\
         & & \checkmark & \checkmark & \checkmark & \checkmark & \textbf{2.48}& \\\midrule
         (+ Server) & \checkmark & \checkmark & \checkmark & \checkmark & \checkmark & \textbf{1.74}& \textbf{86M}  \\
         \bottomrule
    \end{tabular}
    \vspace{1mm}
\end{table}

\textbf{Scalability under various scenarios.}
We have examined the performance of CLIP ViT-B/16 by varying the different available devices and different requesters that have input data and initiate and request the inference. As shown in \reftab{device-availability}, cloud-based inference and our split inference show similar performance, but our split architecture has an advantage in resource usage. 
Interestingly, if we have a powerful GPU-equipped device, \emph{\ours} (with a server) can perform the inference faster than cloud computing, since we benefit from both powerful resources and parallel processing. Furthermore, we initiated the inference across different devices and it showed a similar inference time of 2.47s to 2.51s using \emph{\ours}. Also, our distributed deployment is over neighboring devices, and thereby the latency does not affect the inference latency much.

\subsection{Share Architecture in Multi-Task Inference}\label{sec:eval-share}

On top of the split architecture of multi-modal models, we examined our shared architecture in multi-task scenarios.

\begin{table}[!t]
    \centering
    \caption{Deployment cost and latency when four requests from four different tasks are initiated at the same time. Our split-and-share architecture only deploys one each for the vision encoder of ViT-B/16 on desktop, text encoder with CLIP TRF on laptop, and audio encoder on Jetson, respectively.}
    \label{tab:tradeoff-share}
    \setlength{\tabcolsep}{2pt}
    \begin{tabular}{lllllccc}
        \toprule
        Task& \multicolumn{4}{c}{Total \# of Param} & \multicolumn{2}{c}{Latency (sec)}  \\ \cmidrule(lr){2-5}\cmidrule{6-7}
         & \multicolumn{2}{c}{w/o Sharing} & \multicolumn{2}{c}{w/ Sharing} & w/o Sharing & w/ Sharing \\ \midrule
        
        Retrieval && 124M && \textbf{124M}& 2.48 & \textbf{2.48} \\
        + Encoder VQA & (+124M) &248M & (+1K) &\textbf{124M} & 2.48 & \textbf{2.50}\\
        + Alignment & (+209M) &457M & (+85M) &\textbf{209M} & 3.73 & \textbf{4.87} \\
        + Classification & (+86M) &543M & (+52K) &\textbf{209M} & 3.73 & \textbf{4.97}\\
        \bottomrule
    \end{tabular}
    \vspace{-1mm}
\end{table}

\textbf{Q4: Shareable architecture across different tasks.}
We can support more tasks at a low cost by reusing modules. To evaluate our split-and-share architecture under multi-task scenarios, we deployed only one common module for each--vision encoder of ViT-B/16 on desktop, text encoder of CLIP TRF on laptop, and audio encoder of ViT-B on Jetson Nano. Then, we design as the requests from all tasks are initiated at the same time and showed the latency and the deployment cost. We compare with our framework non-sharing modules, which means each task has dedicated modules and does not suffer from any interference from other tasks. As shown in \reftab{tradeoff-share}, by sharing modules, we can highly reduce the deployment cost by up to 61.5\% when deploying four different tasks. More in detail, for the first image-text retrieval, an image encoder and a text encoder are deployed at first. If we add a VQA task, which needs the image and text encoders, we can reuse the ones in the image-text retrieval task.

\begin{implication}[Memory efficiency II]
    We save deployment cost of up to 61.5\% by reusing common modules across tasks.
\end{implication}

\begin{table}[!t]
    \centering
    \caption{Comparison to baselines, where Mega is Megatron-LM. `--' denotes unavailability.}
    \label{tab:comparison-result}
    \setlength{\tabcolsep}{3.5pt}
    \begin{tabular}{lllllllll}
        \toprule
         & \multicolumn{4}{c}{Latency (sec)} & \multicolumn{2}{c}{\# Param} \\\cmidrule(lr){2-5} \cmidrule{6-7}
         & Optimus & DistMM & Mega & \emph{\ours} & Mega & \emph{\ours} \\\midrule
        VQA & 1.57 & -- & 2.71 & 2.71 & 1.2B & 1.2B \\
        Retrieval & -- & 2.48 & 3.03 & 2.48 & 124M & 124M \\
        Alignment & -- & -- & 0.99 & 0.55 & 209M & 209M \\\midrule
        Retrieval+Alignment & -- & -- & 3.03 & 2.80 & 333M & 209M \\
        \bottomrule
    \end{tabular}
    \vspace{-4mm}
\end{table}

\textbf{Comparison.} 
Lastly, we compare the latency and total memory usage with baselines. As shown in \reftab{comparison-result}, Optimus and DistMM achieve better latency than ours due to their tensor parallelism.\footnote{Existing multi-modal training approaches are not open-source, so we estimate the computation time with (tensor) parallelism as the ideal performance, proportionally reduced based on the number of devices.} However, they require frequent exchanges of intermediate output between partitioned modules, making the latency highly affected by network conditions. Furthermore, they are only designed for one specific task and cannot be used in other tasks. On the other hand, we evaluate Megatron-LM by applying model parallelism for each functional module. However, it cannot benefit from parallel processing across encoders, resulting in higher latency. Also, in multi-task settings, these conventional approaches consume more memory due to their inability to share modules across tasks.

\subsection{Discussions}\label{sec:eval-discussion}
\footnotetext[4]{For example, using LLaVA-Next-7B on Tesla L40S, inference time for batch sizes of 1, 10, and 20 are 1.28s, 4.90s, and 9.16s, respectively.}

\textbf{Non-parallelizable models.} 
Although we achieve a good deployment cost and inference time while maintaining the accuracy, we do not benefit from parallel processing on non-parallelizable architectures. For example, a language model used in VQA or image captioning, e.g., LLaVA~\cite{liu2023visual}, is the most compute-intensive part and is not parallelizable, as they are task heads after the encodings. However, as we pointed out, our architecture is orthogonal to the model compression or model partitioning on LLM, and we can reduce the inference time by applying lightweight models such as Phi-3-mini or TinyLlama-1.1B or a partitioned architecture.

\textbf{Multiple requests.} 
By reusing existing modules, there adversely comes a queuing delay in processing requests, where the next request has to wait until the previous request finishes on the shared module. Thereby, as described in \reftab{tradeoff-share}, the inference latency in module sharing is slightly increased from 3.73s to 4.97s due to sequential inference on the shared module. If we do not share modules, we can prevent queuing delays. However, the common modules to redundantly deploy keep increasing proportional to tasks, while \emph{\ours} is proportional to the number of modalities and tasks. If we use the same modalities across various tasks, we can reuse the modules again and again. Furthermore, this overhead happens not only in sharing architecture but also in non-sharing architecture with multiple requests. This queuing concern can be solved by \emph{batch inference} of multiple requests at the module level, by aggregating requests--either from the same task or from different tasks but sharing the same module. For example, we can group all the images that will be injected into the same vision encoder and process them at once. Similarly, multiple requests from the same task can be processed as a batch (with a slightly longer encoding time\footnotemark[4] than single-request inference).

\textbf{Dynamic network conditions.}
Although network connectivity is dynamic due to network fluctuation, our experimental results have shown that communication latency is negligible compared to computation time. Therefore, short-term network variations have a minimal impact on overall inference latency. Regarding long-term changes such as device availability, \emph{\ours} can provide reallocation with some switching costs. These switching and relocation overheads can be further optimized through adaptive placement.

%% file: 7_conclusion_v3.tex
We proposed \emph{\ours}, a novel distributed framework for on-device multi-modal inference across multiple tasks. Our approach allows users to \emph{plug} in the pretrained models and \emph{play} them for zero-shot inference by deploying models across multiple resource-constrained devices. We first \emph{split} the multi-modal models into modality-wise encoding modules and a task-specific module without modification and then \emph{share} the common modules across different tasks, reducing the resource requirements. 
To address the cross-model dependencies due to module sharing, we provide a \emph{module-level} placement along with \emph{per-request parallel} routing to optimize the inference latency.
We validated our framework using 14 models across 10 benchmarks and five tasks that \emph{\ours} leads to reduced memory usage and latency while maintaining accuracy. 

In future work, we will address challenges in non-parallelizable foundation model architectures by developing more split and more shared techniques to further reduce the deployment cost and inference time. Furthermore, while we achieved optimal placement in most cases, our greedy solution becomes more non-trivial depending on the number and capacity of devices, as well as the number of models, requests, and tasks. {Additionally, although we primarily use the inference time as our evaluation metric--since the inference typically consumes less power than training--the power consumption is still one of the key factors for the battery life of edge devices.} Therefore, to further enhance the system under these various factors, we plan to further optimize the placement and routing problem.
We believe this work contributes to opening and advancing the field of distributed frameworks for edge-only inference in multi-modal multi-task models.